\documentclass{PoS}

\title{Simulations for the JEM--EUSO program with ESAF}

\ShortTitle{Simulations for the JEM--EUSO program with ESAF}

\author{\speaker{Francesco Fenu} for the JEM-EUSO Collaboration\footnote{for collaboration list see PoS(ICRC2019)1177} \\ \\
        Universit\`a degli studi di Torino, INFN Torino, Italy\\
        E-mail: \email{francesco.fenu@gmail.com}}
\author{Kenji Shinozaki, Hiroko Miyamoto, Alessandro Liberatore \\
        Universit\`a degli studi di Torino, INFN Torino, Italy}
\author{Naoto Sakaki \\
        RIKEN Advanced Science Institute, Wako, Japan}
\author{Sergei Sharakin, Mikhail Zotov \\
  Skobeltsyn Institute of Nuclear Physics, Lomonosov Moscow State University, Russia}
\author{Gabriel Chiritoi \\
  Space Science Institute, Magurele, Romania}

\abstract{JEM--EUSO is an international program for the development of space based Ultra High Cosmic Ray
observatories. The program consists of a series of missions which are either
under development or in the data analysis phase. The instruments consist of wide field of view
telescopes, which operate in the UV range, designed to detect the fluorescence light emitted by
extended air showers in the atmosphere. We describe the simulation software ESAF
and show the physical assumptions done in it. We present here the implementation of the POEMMA, K--EUSO,
TUS, Mini--EUSO, EUSO--SPB and EUSO--TA detectors in the simulation software ESAF.}

\FullConference{36th International Cosmic Ray Conference -ICRC2019-\\
		July 24th - August 1st, 2019\\
		Madison, WI, U.S.A.}

\begin{document}

\section{Introduction}
The Joint Experiment Mission--EUSO (JEM--EUSO) program is an international effort devoted to the study of Ultra High Energy Cosmic Rays (UHECR).
The program consists of a series of missions both in preparation or already performed. All such detectors demand an extensive simulation work to estimate the performances and to support the analysis of the events.

The Euso Simulation and Analysis Framework (ESAF) is the simulation and analysis software developed in the framework of the JEM--EUSO program \cite{berat}.
The detector configurations, orbital altitude, atmospheric condition, electronics and several kind of physical events can be simulated.
The ESAF software performs the simulation of the UHECR event, its detection and the shower parameter reconstruction.
The software, which consists of $\sim 2 \cdot 10^{5}$ lines, is written in C++,  following an object oriented approach and makes use of the
\textit{root} package of CERN.
The ESAF software is described here and few examples of its functionalities are shown.

\section{The simulation framework}
The \textit{Simu} framework of ESAF performs the simulation of the entire physical process from shower to telemetry.
Several air shower generators are currently implemented like
SLAST, a parametrical simulator developed in the framework of the JEM--EUSO collaboration, CONEX and CORSIKA \cite{CORSIKA}.
An atmospheric modelization according to the 1976 Standard US Atmosphere \cite{US_Atmosphere} is implemented,
as well as the fluorescence yield parametrization \cite{nagano} and the standard Cherenkov production theory.
The LOWTRAN 7 \cite{lowtran7} package is embedded into ESAF to simulate the atmospheric transmission.
Both Rayleigh scattering and Ozone absorption are taken into account.
The presence of clouds is simulated in parametrical way as a uniform layer with a predefined optical depth,
altitude and thickness.
The photons spectral distribution, timing and direction are produced in the so called \textit{BunchRadiativeTransfer}.
Depending on the solid angle covered by the telescope pupil a number of
single photons is produced and propagated to the detector position.

The optics simulation is performed through the simulator developed in RIKEN \cite{takki} or a parametrical simulator.
In the latter case, the position of the spot on the focal surface is parametrized in analytical way as a function of the entrance angle.
An additional random component of Gaussian shape is added to parametrize the point spread function and an efficiency factor
takes into account the transmittance of the lenses.

The focal surface structure is read in by parameter file, where the position and orientation of each single PMT is defined.
The overall detector efficiency, as product of quantum and collection efficiency,
can be parametrized at the single pixel level. The PMTs have an average gain set by parameter and the front end
electronics is treated in a simplified way, with a threshold on the current pulse delivered by the PMT.
The background is added at the level of front end electronics and can be also set at the pixel level.
The electronics simulation is then concluded by the trigger.
Several algorithms \cite{bertaAlgorithms} have been therefore implemented into ESAF and can be used in combination.

\section{The reconstruction framework}
\begin{figure}[h!]
\centering
\includegraphics[width=0.8\textwidth]{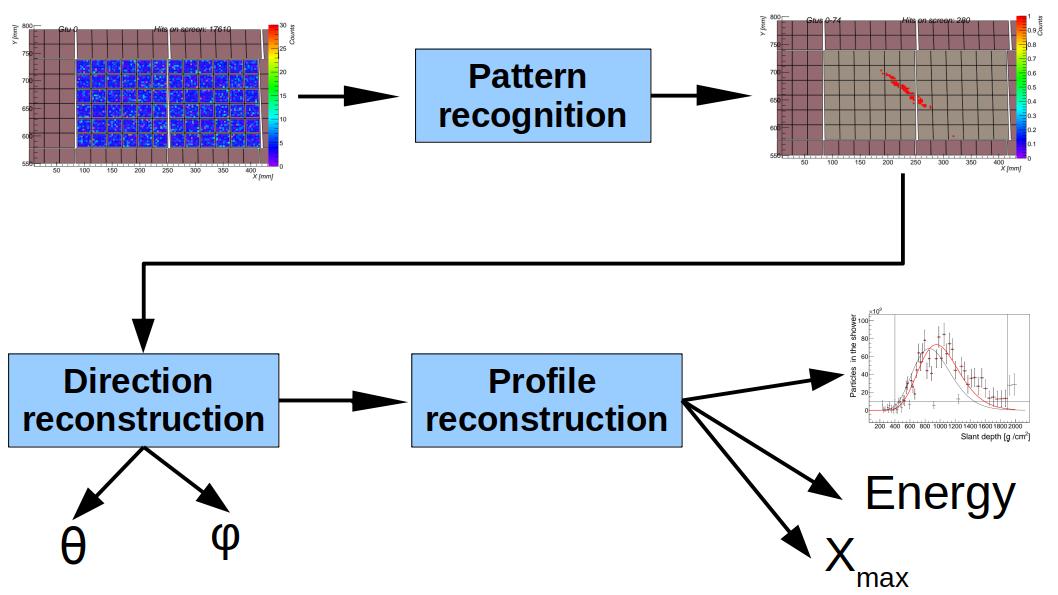}
\caption{the general scheme of the reconstruction framework.}
\label{fig:RecoScheme}
\end{figure}

The first task to accomplish in the reconstruction phase (see Fig. \ref{fig:RecoScheme}) is to recognize the signal from the shower into the
transmitted data. A series of clustering algorithms have been implemented in the code.
The \textit{PWISE} algorithm \cite{guzman} searches for high concentrations of the signal on the single pixels.
The algorithm selects therefore pixels--GTUs whose signal is above a certain threshold and checks whether this signal excess is persistent over time.
The \textit{LTTPatternRecognition} is modeled following  the level 2 trigger philosophy of the JEM--EUSO project \cite{bayer}. The integration of the signal is performed
in a set of predefined test directions and the direction which maximizes the integral is the one chosen to reconstruct the event.
Both algorithms exploit the morphology of the signal, which can be seen as a spot of light moving on the focal surface at the projected speed of light.

The \textit{TrackDirection2} is the angular reconstruction module \cite{mernik}.
Several algorithms are implemented in it but two main families of algorithms are in use: analytical and numerical.
In the first group, a fit on the speed of the shower is performed to determine its inclination after the Track Detector Plane (TDP) has been identified.
The algorithms can be used in an iterative way to improve the knowledge of the arrival direction of the shower.
In the numerical approach a series of test geometries is chosen and the deviation with respect to the timing and arrival direction of the measured event is calculated.
The test direction that better describes the measured event properties is the one that delivers the arrival direction.

The energy reconstruction is performed in the \textit{PmtToShowerReco} module \cite{fenu}. The first step consists in an
estimate of the position of the shower maximum in the atmosphere. Two methods can be used depending on whether the Cherenkov reflection feature is detected.
In case a reflection feature is present, the geometry can be constrained very well thanks to the timing difference of maximum and Cherenkov peak.
If the peak is not present, an iterative procedure must be started to identify a volume in the space which is likely to contain the maximum.
As initial step, an assumption on the shower maximum position must be made and a scan in the parameter space is then required to test the compatibility of the
resulting shower profile with the measured one.

The count profile of the shower is reconstructed by selecting a collection area that follows the cluster of pixels.
The size of this selection area is a trade--off between
the need to collect the highest possible fraction of the signal and the need to limit the background.
The detector modeling allows then to correct for the detector efficiency and to retrieve a photon curve at the entrance pupil.
The reconstruction of the geometry is fundamental to apply the correction of the atmospheric transmittance. As a final step, the
parametrization of the energy distribution of the secondary particles allows to calculate the fluorescence and Cherenkov yield.
All this information allows then to reconstruct the the secondary particle profile of the shower.
The reconstruction of the energy and $\mathrm X_{max}$ is then performed through a fit with a shower profile function to the reconstructed shower profile.

\section{The implementation of the JEM-EUSO program missions}
The complexity of the JEM--EUSO program requires an extensive effort to perform the study of all the different detectors performances.
The detectors are either space based, like Mini--EUSO \cite{mini-EUSO}, TUS \cite{TUS}, K--EUSO \cite{K-EUSO} and POEMMA \cite{POEMMA} or balloon based, like EUSO--SPB1 \cite{EUSO-SPB}. EUSO--TA \cite{EUSO-TA} is the only one simulated on ground.
All the detectors are in nadir mode, except for EUSO--TA which is looking in direction north west with an elevation of 15 degrees.
All the detectors have a different focal surface: Mini--EUSO, EUSO--TA and EUSO--SPB1 are single PDMs detectors while POEMMA and K--EUSO have a multi--PDM layout.
TUS on the other hand, consists of an array of 256 PMTs on a square of 16 $\times$ 16 side.
The time frame is always of 2.5 $\mu$s with the only exception of TUS, which has a frame of 0.8 $\mu$s.

\begin{figure}[h!]
\centering
\includegraphics[height=4.5cm]{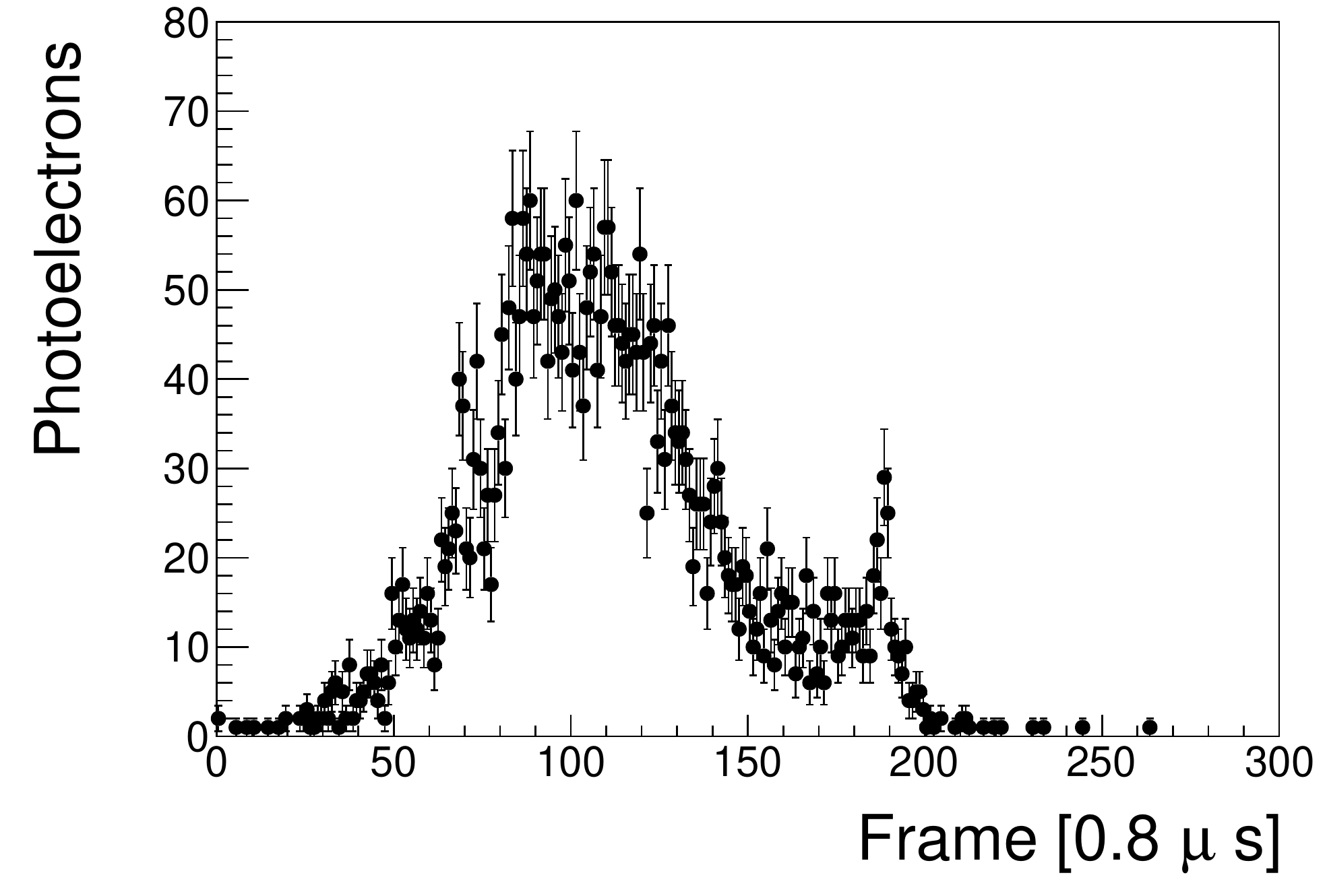}\hfill
\includegraphics[height=4.55cm]{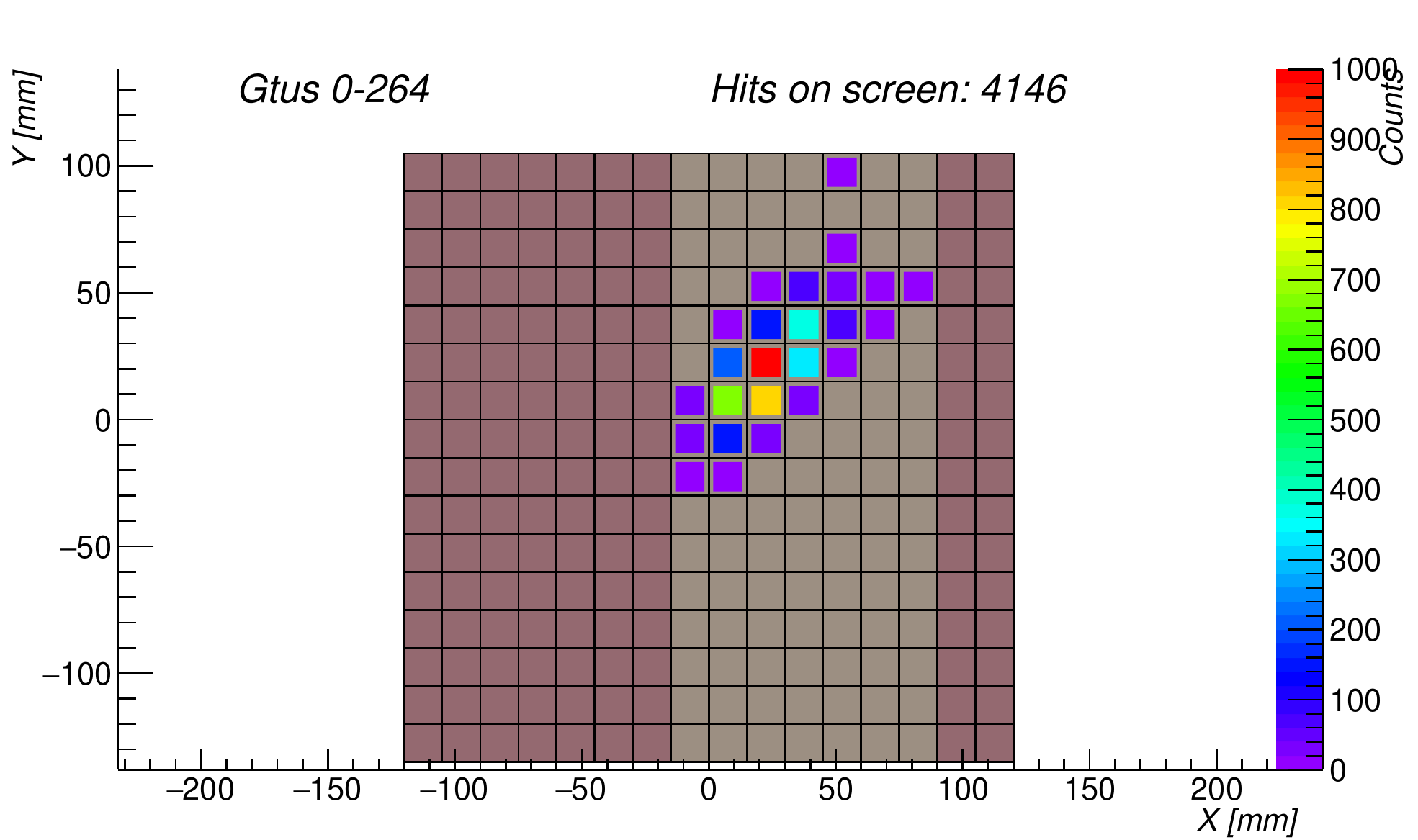}
\caption{a $10^{21}$ eV, 60 degrees zenith angle event. On the left: the photoelectron profile for the TUS detector. On the right: the photoelectron image for TUS.}
\label{fig:TUS}
\end{figure}

\begin{figure}[h!]
\centering
\includegraphics[height=4.5cm]{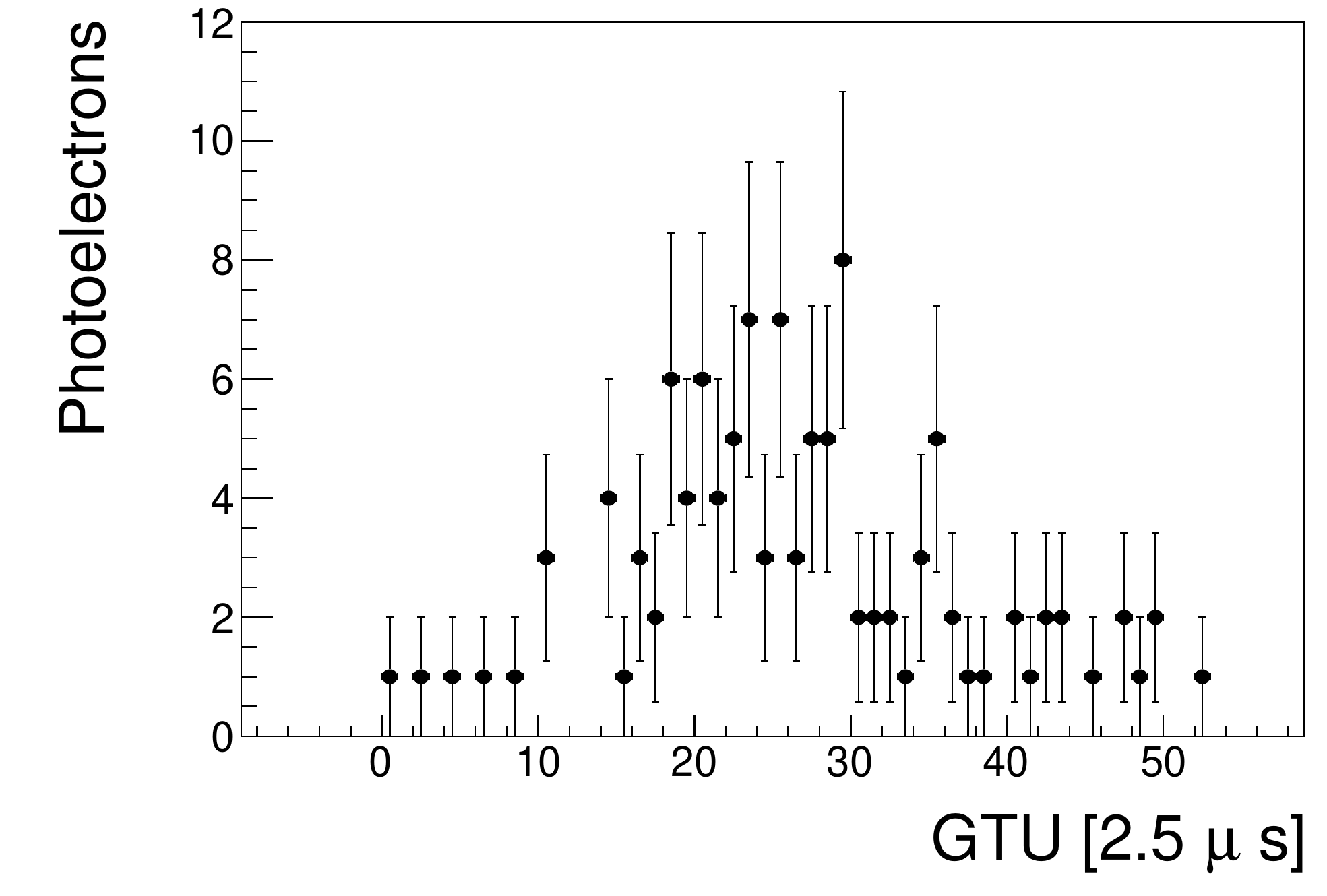}\hfill
\includegraphics[height=4.55cm]{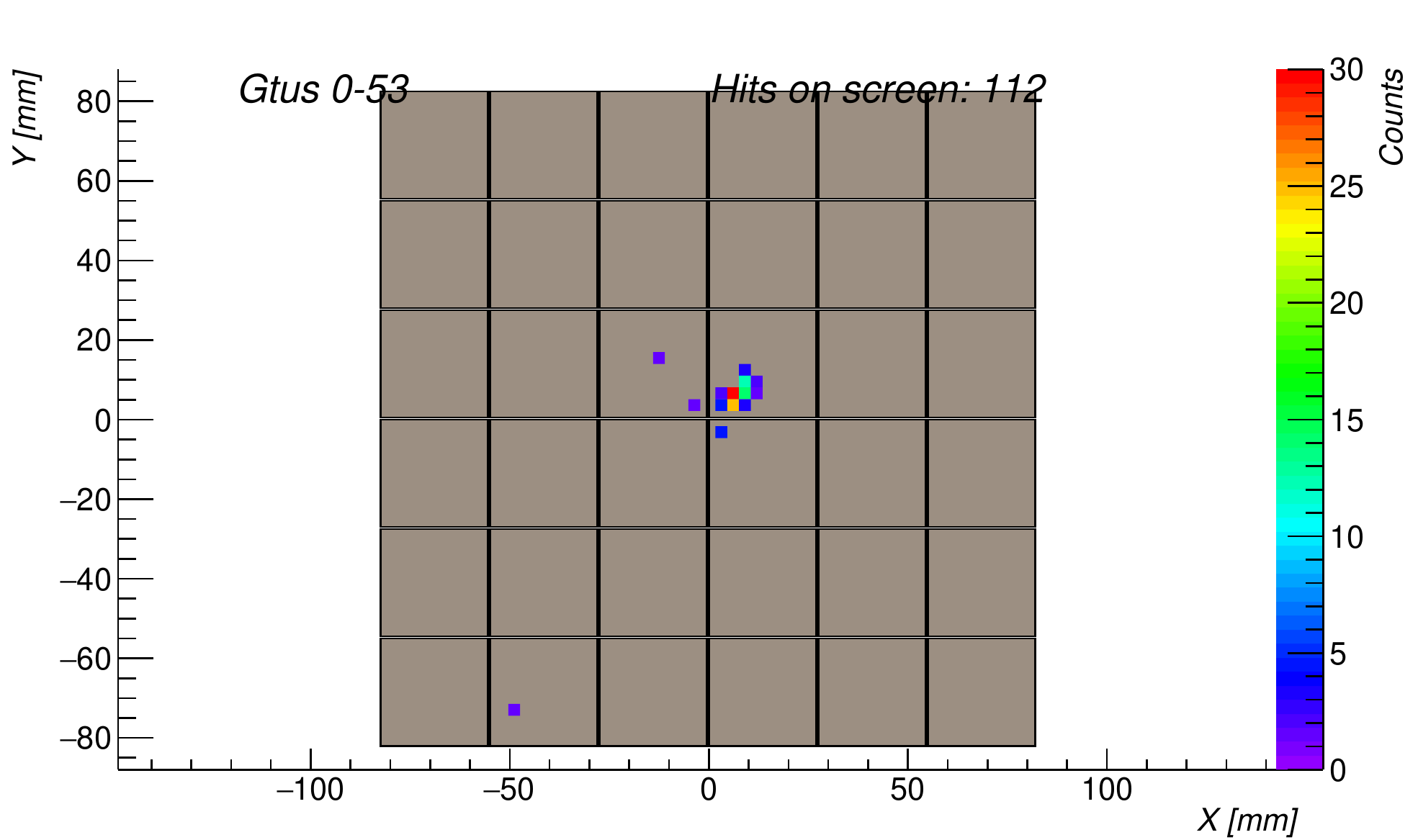}
\caption{a $10^{21}$ eV, 60 degrees zenith angle event. On the left: the photoelectron profile for the Mini--EUSO detector. On the right: the photoelectron image for Mini--EUSO.}
\label{fig:Mini-EUSO}
\end{figure}

In Fig. \ref{fig:TUS} and \ref{fig:Mini-EUSO} the signal of a $10^{21}$ eV 60 degrees zenith angle shower is shown for the TUS and Mini--EUSO detector respectively. 
As it is very clear, Mini--EUSO is at the detection threshold while TUS would have enough sensitivity to detect such events.
The choice of such a high energy is done only for having high enough signal to allow the detection by Mini--EUSO (which is not designed to detect CR).
The signal is shown in terms of photoelectrons to
be able to directly compare the two detectors without considering the substantial differences in the front end electronics.

\begin{figure}[h!]
\centering
\includegraphics[height=4.5cm]{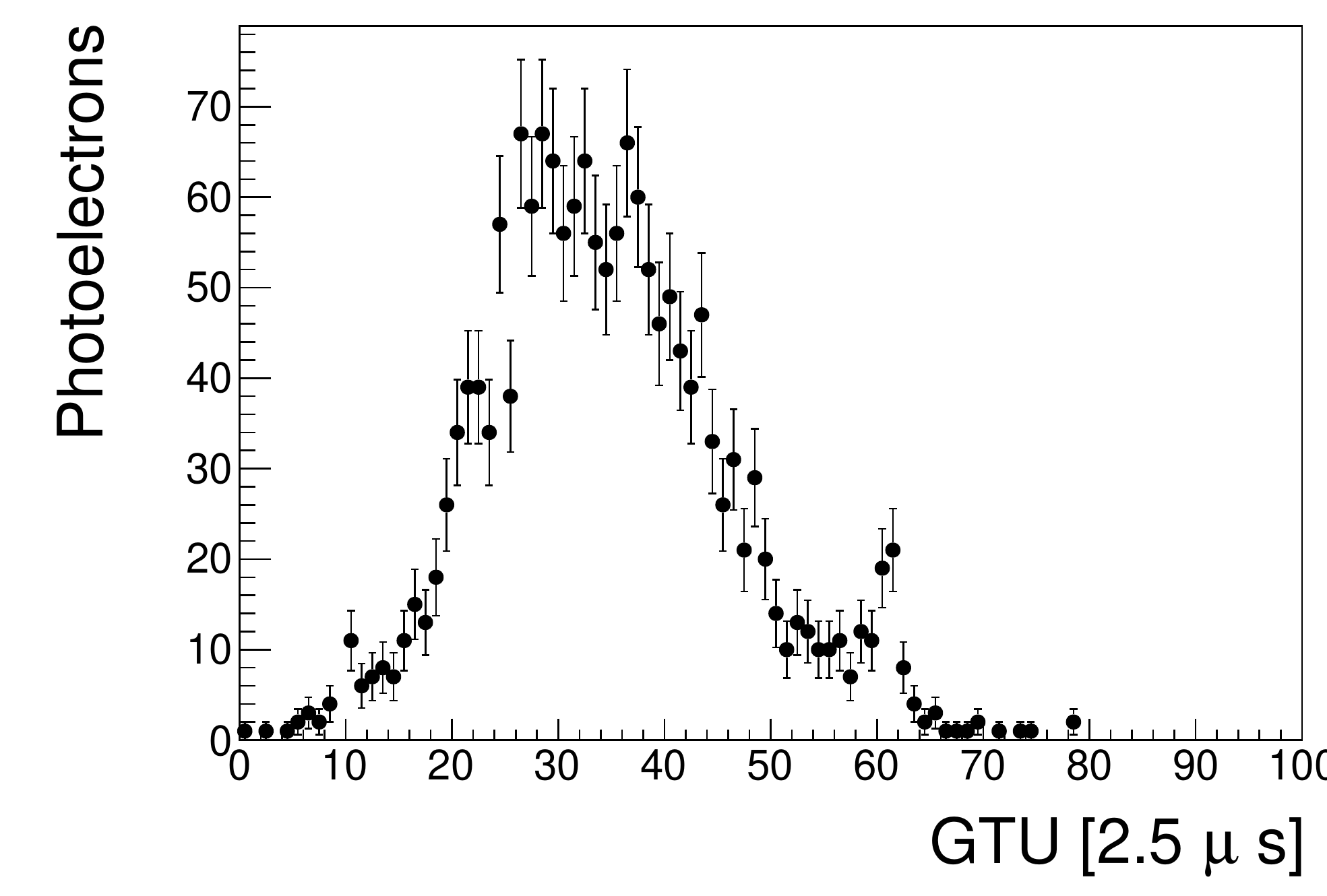}\hfill
\includegraphics[height=4.55cm]{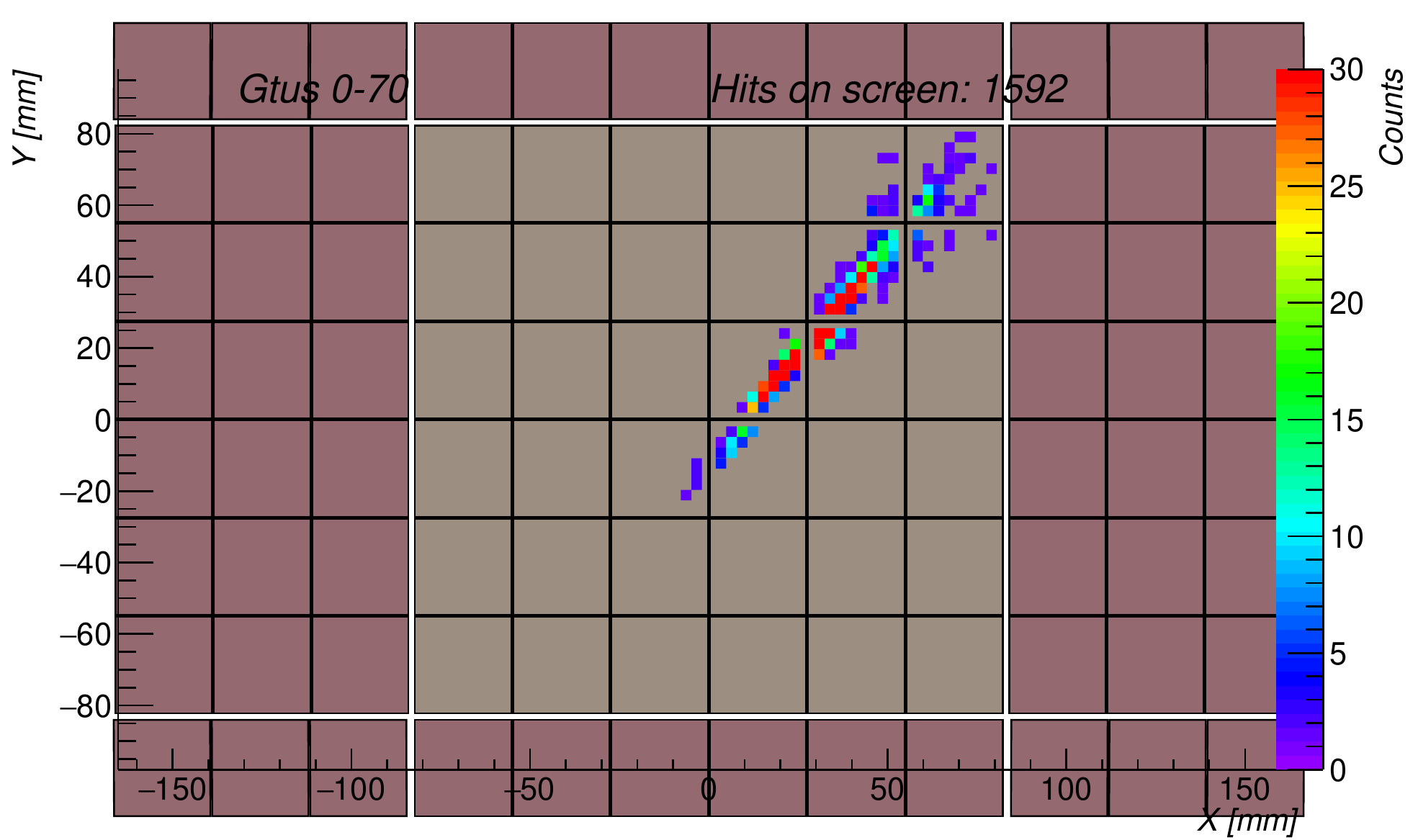}
\caption{a $10^{20}$ eV, 60 degrees zenith angle event. On the left: the photoelectron profile for the POEMMA detector. On the right: the photoelectron image for POEMMA.}
\label{fig:POEMMA}
\end{figure}

\begin{figure}[h!]
\centering
\includegraphics[height=4.5cm]{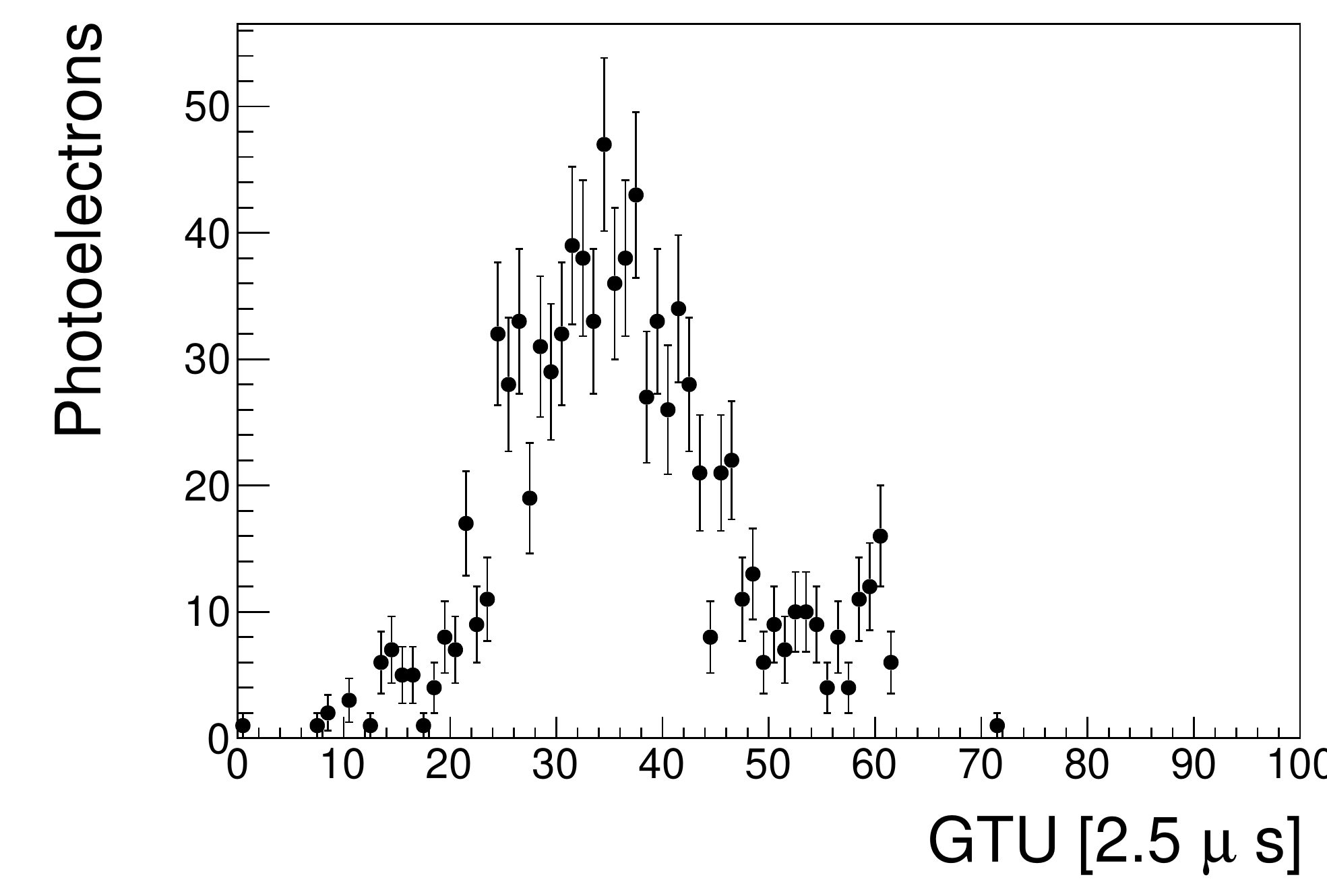}\hfill
\includegraphics[height=4.55cm]{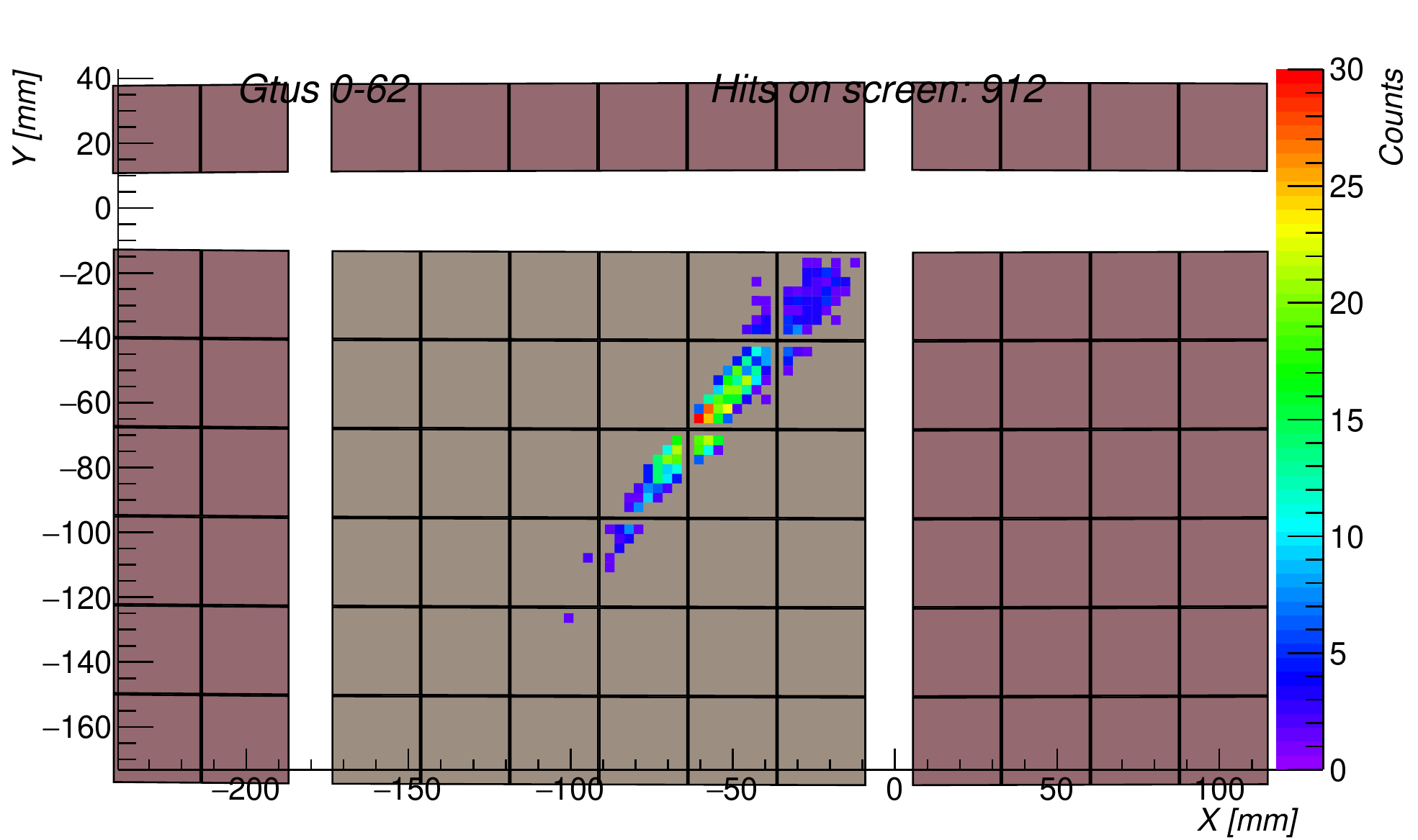}
\caption{a $10^{20}$ eV, 60 degrees zenith angle event. On the left: the photoelectron profile for the K--EUSO detector. On the right: the photoelectron image for K--EUSO. Note: in the new layouts the dead areas between PDMs still visible here are strongly reduced. }
\label{fig:K-EUSO}
\end{figure}

Another example is shown in Figs. \ref{fig:POEMMA} and \ref{fig:K-EUSO} where a a $10^{20}$ eV 60 degrees zenith angle event is shown for the POEMMA and K--EUSO respectively.
Both implementations are still in a preliminary stage and are in evolution but can be still used to give an estimate of the performances.

\begin{figure}[h!]
\centering
\includegraphics[height=4.5cm]{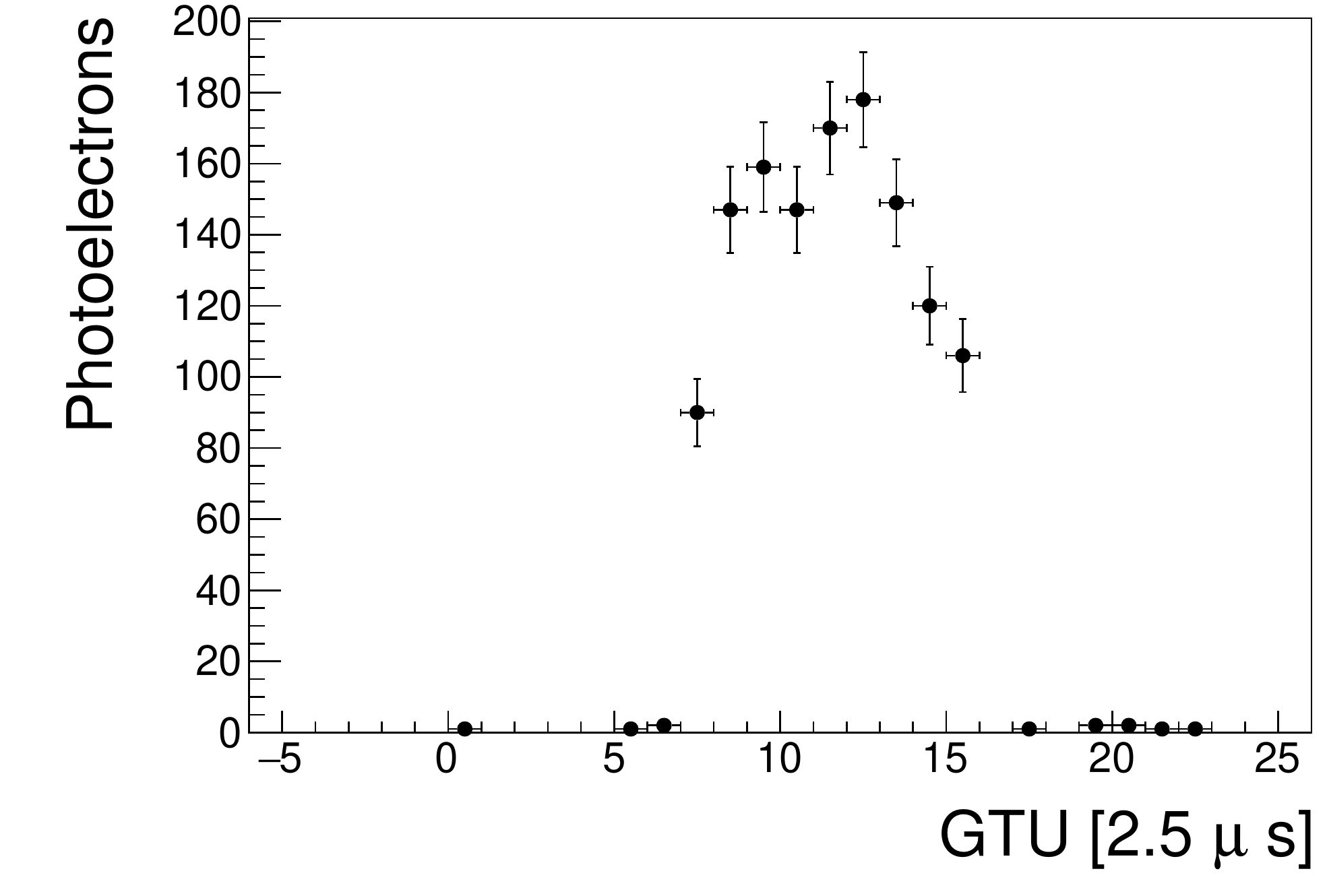}\hfill
\includegraphics[height=4.55cm]{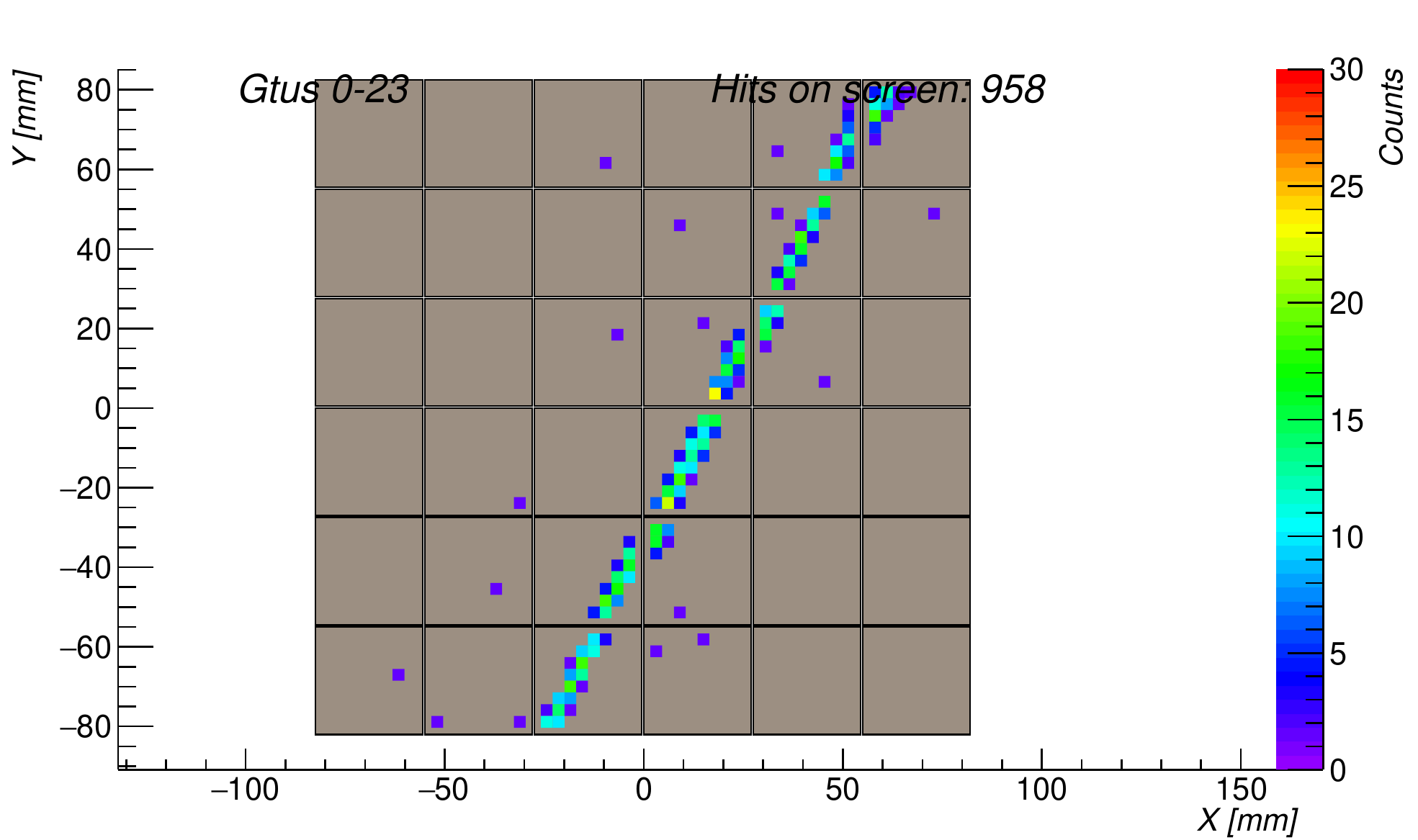}
\caption{a $10^{19}$ eV, 45 degrees zenith angle event impacting 25 km away from the detector. On the left: the photoelectron profile for the EUSO--TA detector. On the right: the photoelectron image for EUSO--TA.}
\label{fig:EUSO-TA}
\end{figure}

\begin{figure}[h!]
\centering
\includegraphics[height=4.5cm]{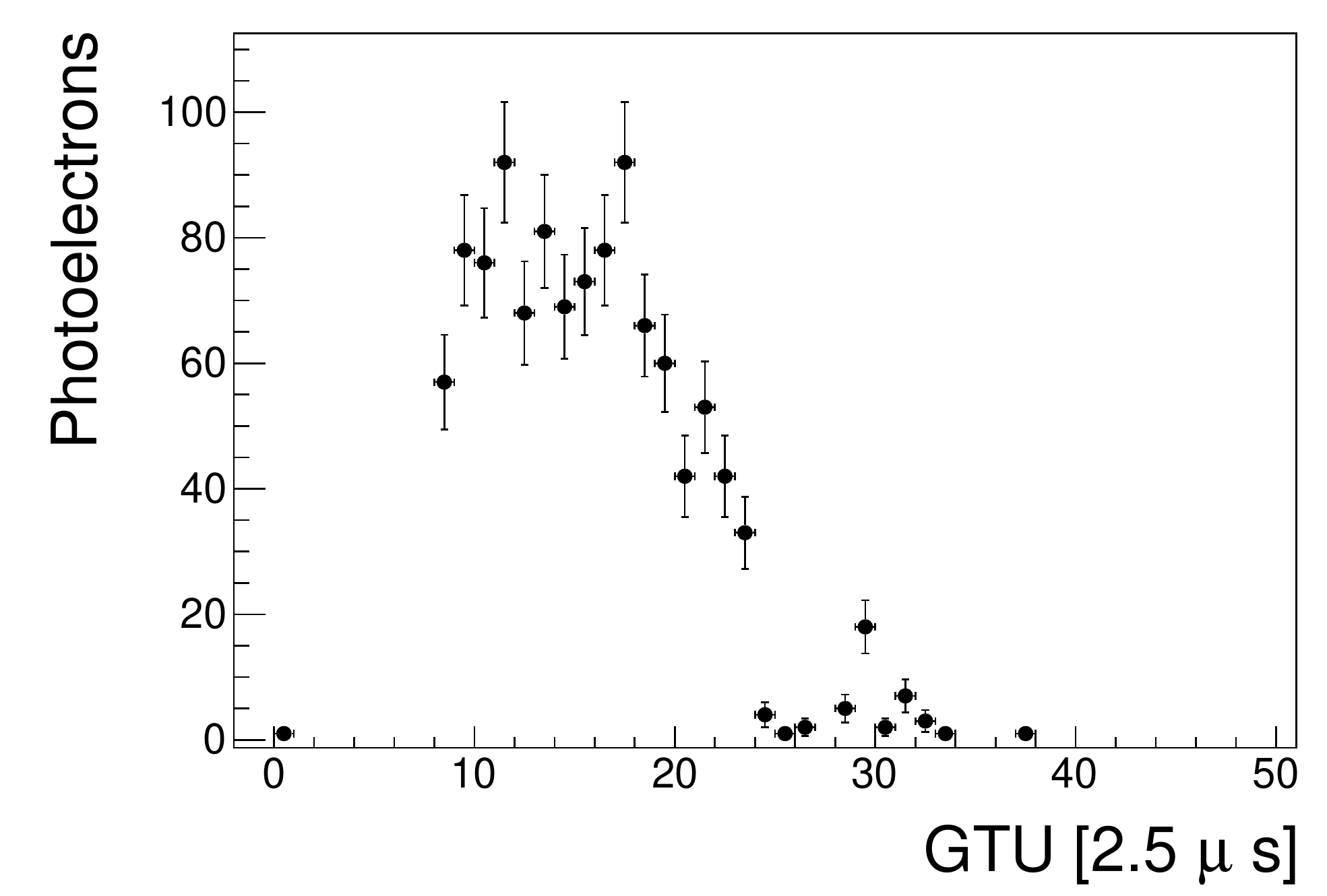}\hfill
\includegraphics[height=4.55cm]{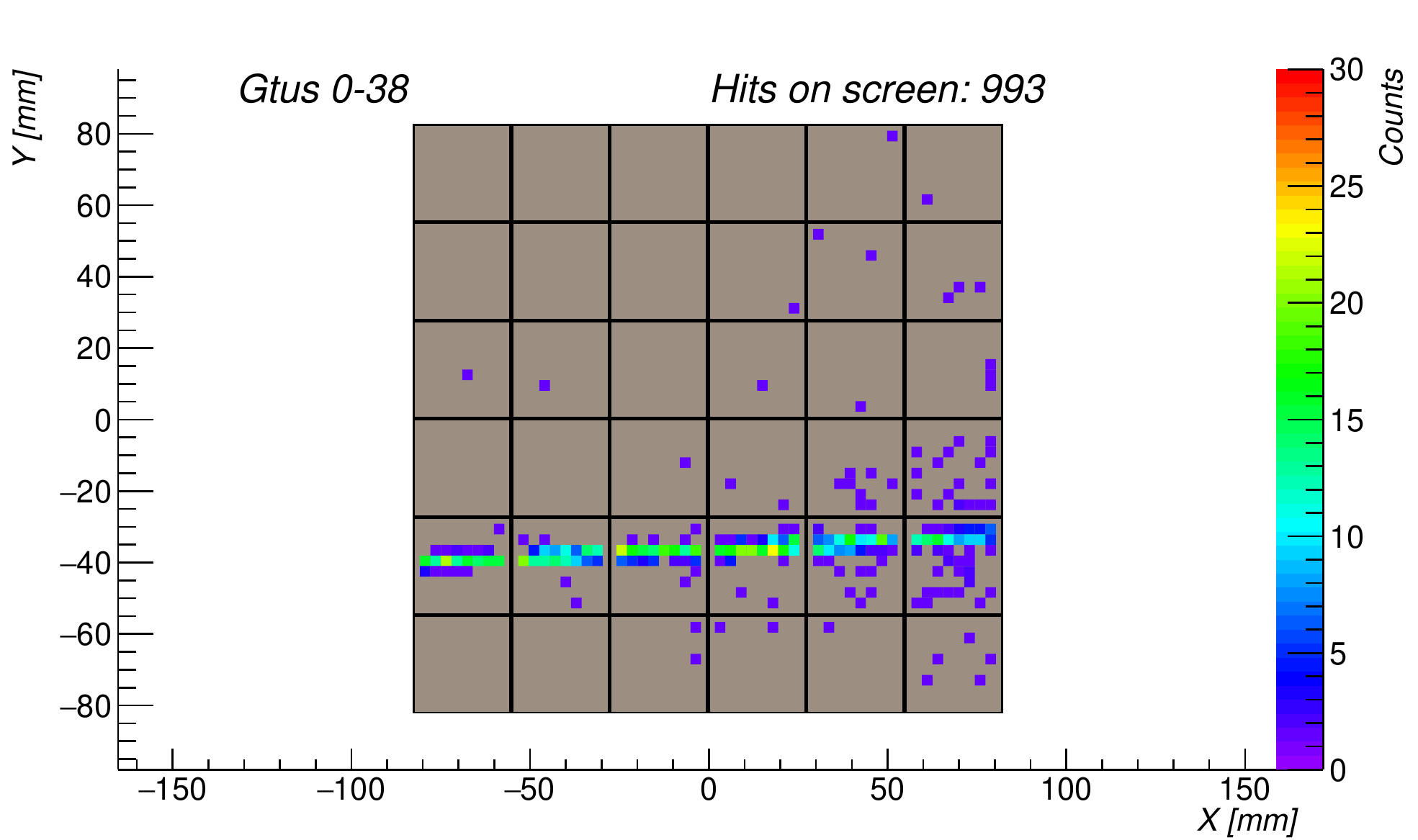}
\caption{a $10^{19}$ eV, 45 degrees zenith angle event. On the left: the photoelectron profile for the EUSO--SPB1 detector. On the right: the photoelectron image for EUSO--SPB1. The detector is at 30 km altitude. }
\label{fig:Ballooon}
\end{figure}

The prototypes from ground and balloon are shown in Figs. \ref{fig:EUSO-TA} and \ref{fig:Ballooon} where we show the EUSO--TA and EUSO--SPB1
implementations.
For both cases (having the detector already being built) we could measure the relative efficiency map at the single pixel level and implement it into ESAF.

\section{The performance of space based UHECR detectors}
An example of the reconstruction functionality of the ESAF framework is shown here.
We show therefore the results for a detector (only as an example) with a 1650 mm pupil radius and a parametrical optics. The detector flies at 550 km altitude and has a 2.5 microseconds GTU.
We assume a background radiance of 500 photons  ns$^{-1}$  m$^{-2}$  sr$^{-1}$ and that the JEM--EUSO first level trigger is in operation.

\begin{figure}[h!]
\centering
\includegraphics[height=5.1cm]{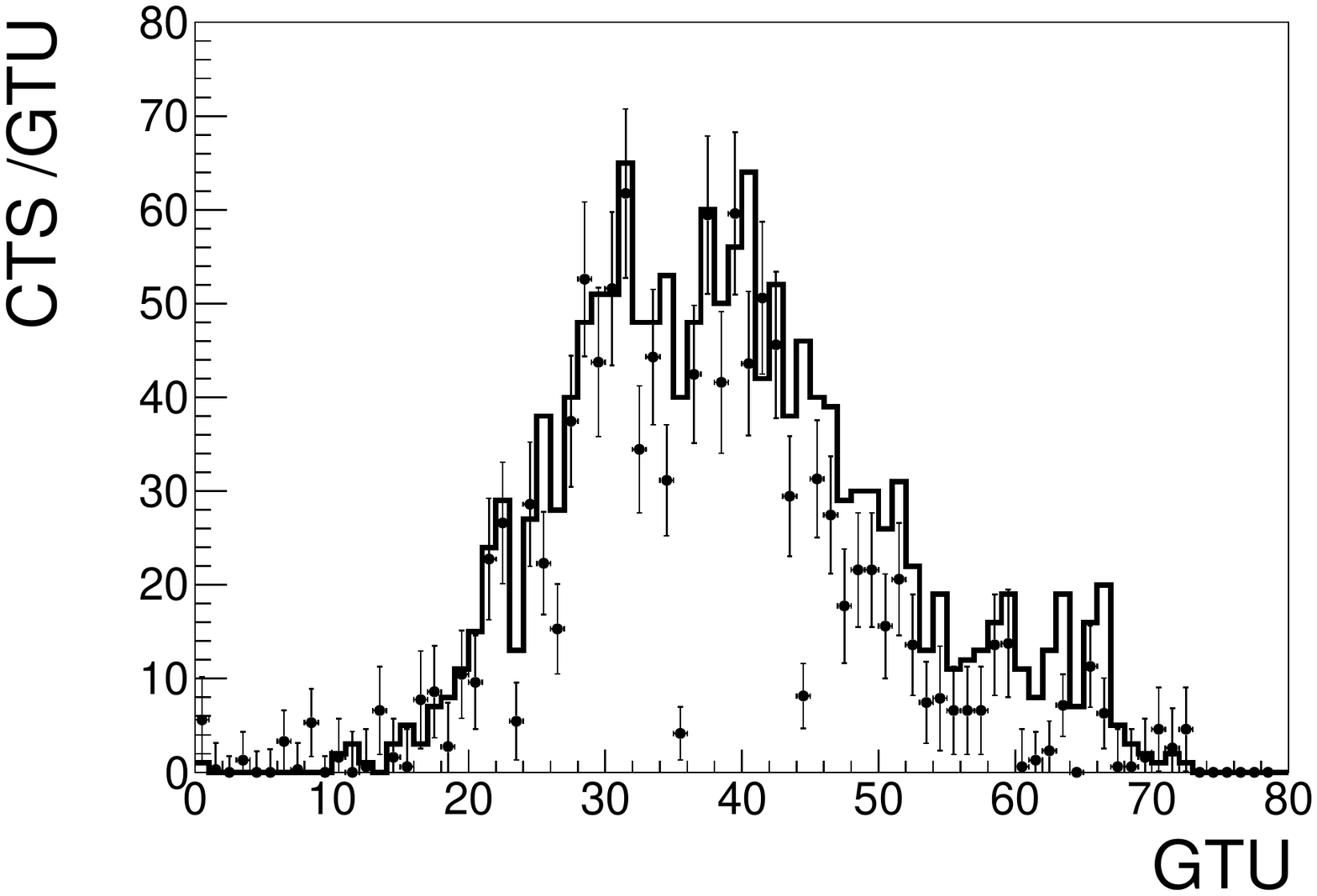}\hfill
\includegraphics[height=5cm]{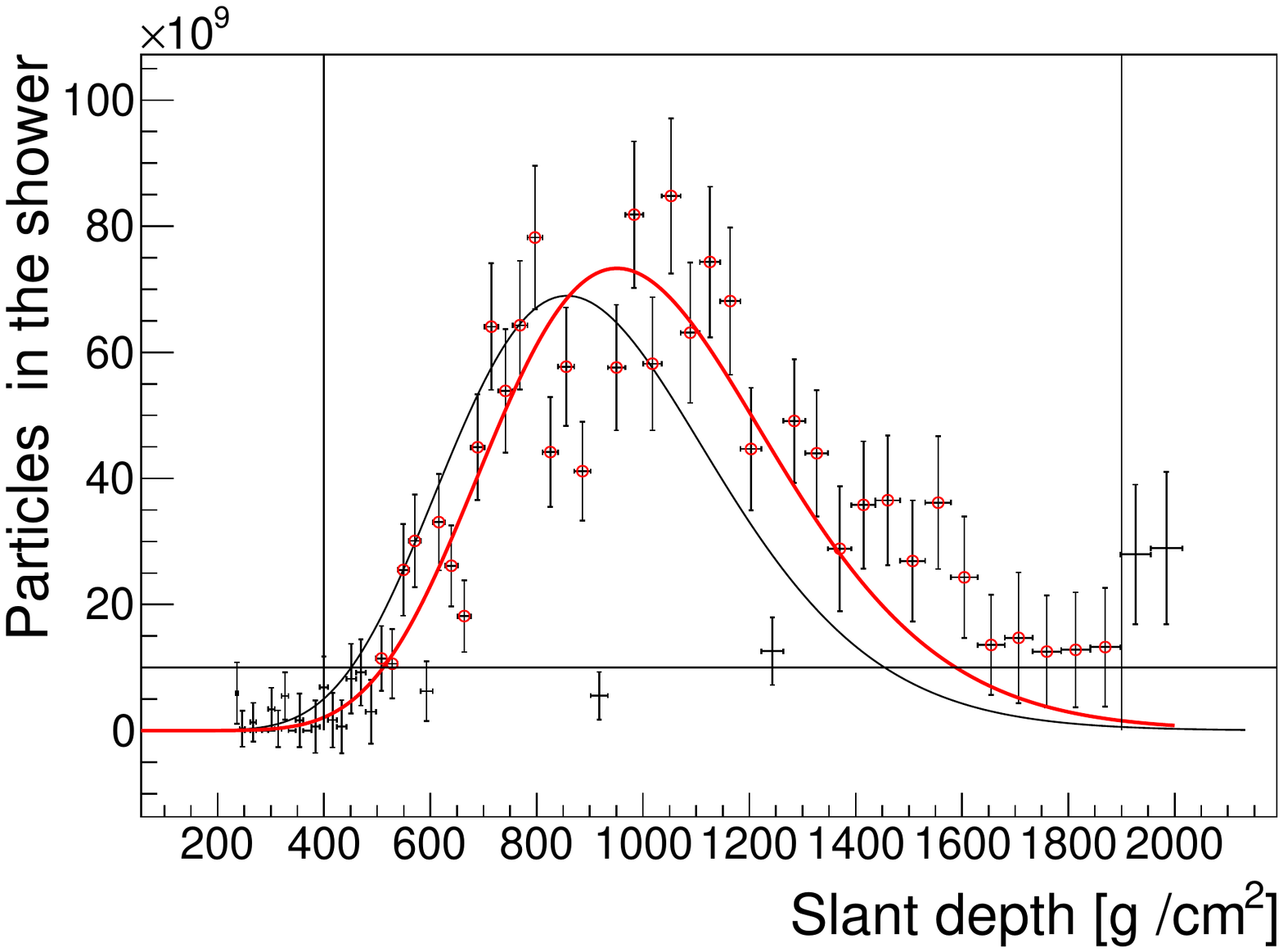}

\caption{a comparison between simulated (line) and reconstructed (points) counts profile on the left. The comparison between simulated (line) and reconstructed shower profile (points) on the right. The fit on the
reconstructed profile is shown in red.}
\label{fig:EventReco}
\end{figure}

In Fig. \ref{fig:EventReco} an example of the reconstruction of the profile is shown.
On the left, the counts profile is reconstructed by using a collection area of 3.5 mm radius and applying the corresponding background subtraction.
The presence of a PMT gap is, for example, visible near the maximum of the profile.
On the right side, the shower profile is shown after the entire reconstruction chain.
A shower parametrization is fit to the reconstructed shower profile and the parameters like energy and $\mathrm X_{max}$ are obtained.
For this shower, of  $10^{20}$ eV, 60 degrees zenith angle, an energy of $1.06 \cdot 10^{20}$ eV is reconstructed. The $\mathrm X_{max}$ has been reconstructed to 951 $\mathrm g / cm^{2}$
(instead of the true 852  $\mathrm g / cm^{2}$).

\begin{figure}[h!]
\centering
\includegraphics[height=4cm]{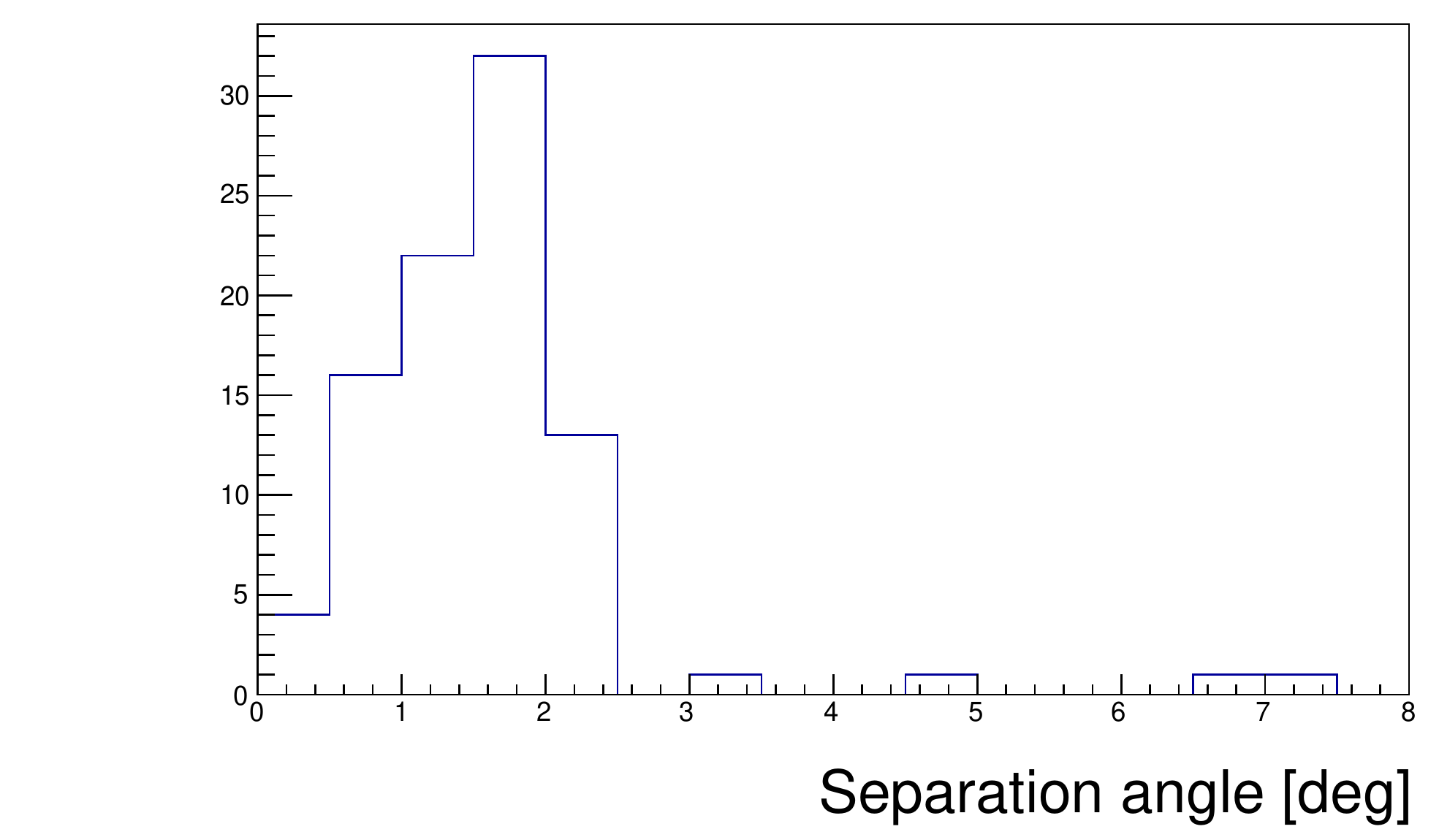}\hfill
\includegraphics[height=4.2cm]{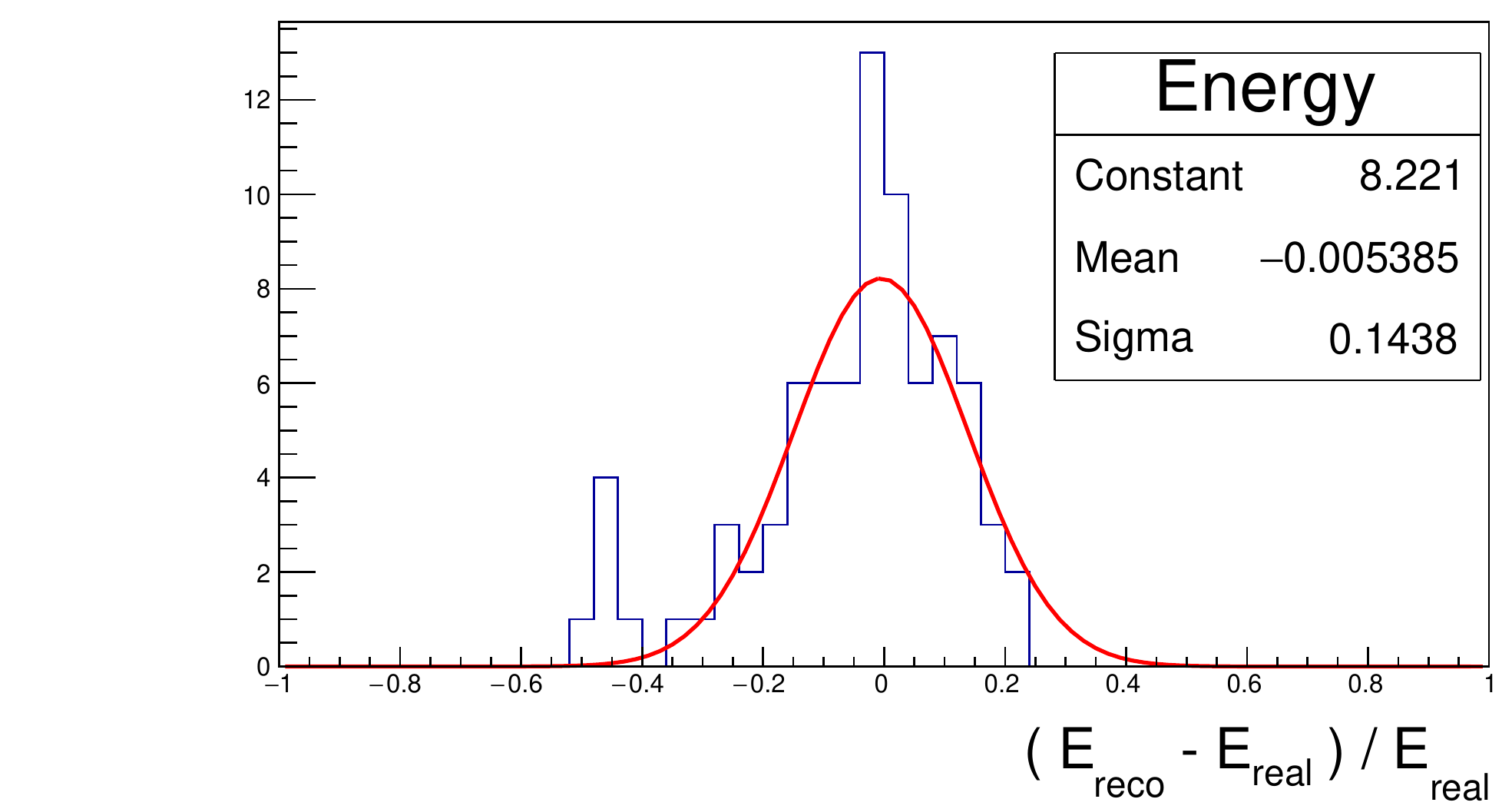}

\caption{the reconstruction performances for $10^{20}$ eV, 60 degrees events in the central region of the FOV. On the left: angular resolution. On the right: energy resolution.}
\label{fig:EventReco_E}
\end{figure}

A minimal sample of 100 events of $10^{20}$ eV, 60 degrees is reconstructed in this configuration. In Fig. \ref{fig:EventReco_E} (left) the angular reconstruction is shown.
The \textit{PWISE} clustering is applied here and the 68$\%$ of the events falls inside $\sim$ 2 degrees. On the right side, the energy reconstruction is shown.
The \textit{LTTPatternRecognition} is used in this case and a resolution of the order of 15$\%$ is therefore achieved. The small outlier group on the left side of this plot is  due
to the gaps between the PDMs.

\begin{figure}[h!]
\centering
\includegraphics[width=0.6\textwidth]{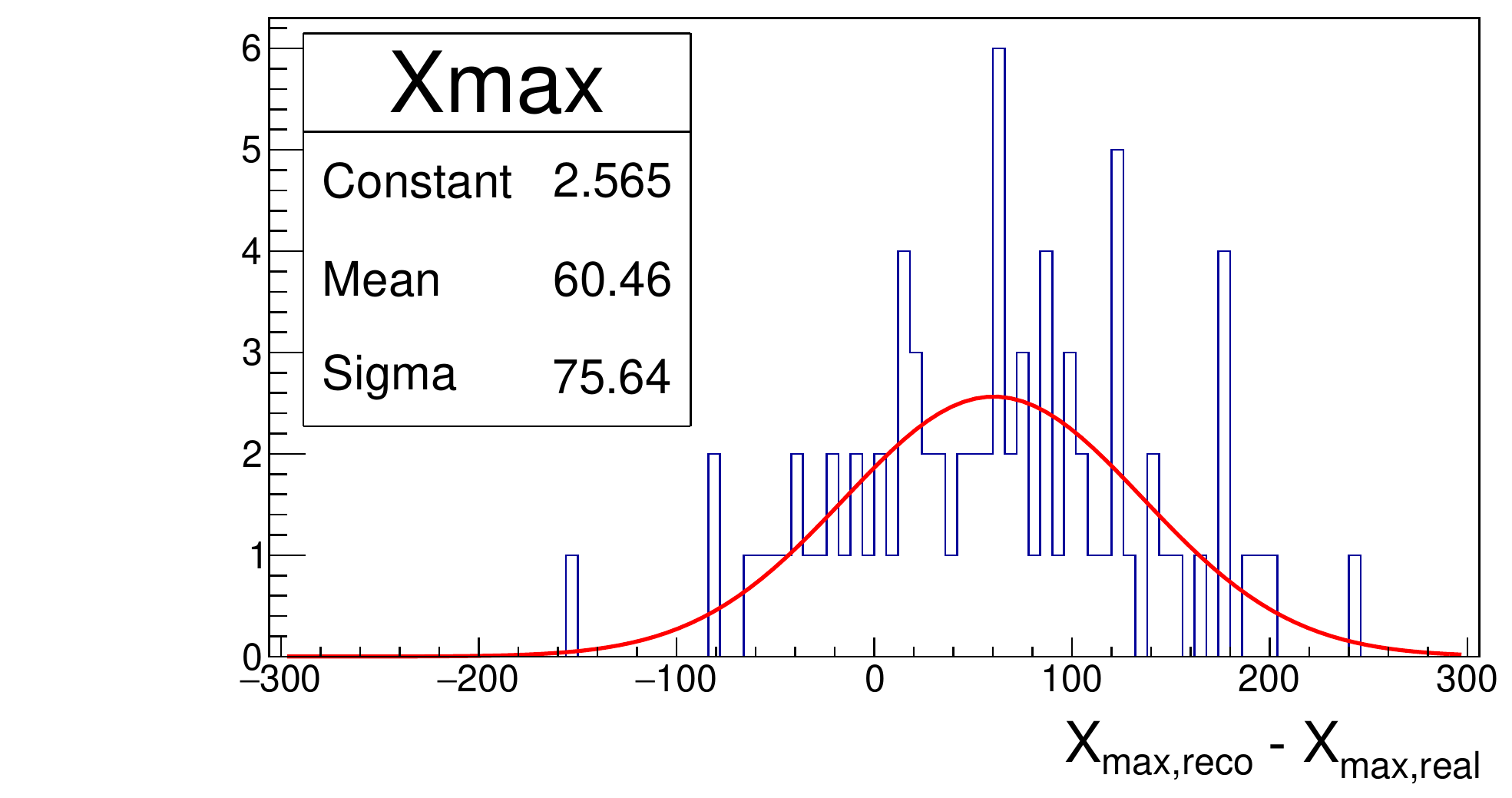}
\caption{the $\mathrm X_{max}$ reconstruction performances for $10^{20}$ eV, 60 degrees events in the central region of the FOV. Only the first iteration of the slant depth method is shown.}
\label{fig:X-MaxResolution}
\end{figure}

To conclude, an example of the $\mathrm X_{max}$ reconstruction is shown here.
The first assumption on the position of the maximum in the atmosphere is done according to the reconstructed shower direction and with an hypothesis on the slant depth of the maximum.
The profile is then reconstructed and the fit on the profile is then converging to the values shown here. We obtain a 70 $\mathrm g / cm^{2}$ resolution.
It must be stressed that this is the first iteration only and a following iterative procedure must be applied to identify a region in the parameter space compatible with the measured profile and to correct the observed bias.

 \section*{Acknowledgments}
 This work was partially supported by Basic Science Interdisciplinary Research
Projects of RIKEN and JSPS KAKENHI Grant (JP17H02905, JP16H02426 and
JP16H16737), by the Italian Ministry of Foreign Affairs and International
Cooperation, by the Italian Space Agency through the ASI INFN agreement
n. 2017-8-H.0 and contract n. 2016-1-U.0, by NASA award 11-APRA-0058 in
the USA, by theDeutsches Zent rum f\"ur Luft- und Raumfahrt,
by the French space agency CNES, the Helmholtz Alliance for Astroparticle Physics funded by the Initiative and Networking Fund of the Helmholtz Association
(Germany), by Slovak Academy of Sciences MVTS JEMEUSO as well as
VEGA grant agency project 2/0132/17, by National Science Centre in Poland
grant (2015/19/N/ST9/03708 and 2017/27/B/ST9/02162), by Mexican funding
agencies PAPIIT-UNAM, CONACyT and the Mexican Space Agency (AEM).
Russia is supported by ROSCOSMOS and the Russian Foundation for
Basic Research Grant No 16-29-13065. Sweden is funded by the Olle Engkvist
Byggm\"astare Foundation.

\end{document}